\documentclass{article}
\pdfoutput=1

\usepackage{arxiv}

\usepackage[utf8]{inputenc} 
\usepackage[T1]{fontenc}    
\usepackage{hyperref}       
\usepackage{url}            
\usepackage{booktabs}       
\usepackage{amsfonts}       
\usepackage{nicefrac}       
\usepackage{microtype}      
\usepackage{graphicx}
\usepackage{subcaption}
\usepackage{array,calc}

\newlength\lena \newlength\lenb \newlength\lenc \newlength\lend
\newcolumntype{P}[1]{>{\centering\arraybackslash}p{#1}} 
\newcommand\mcii[1]{\multicolumn{10}{P{\lenb}|}{#1}}  
\newcommand\mciii[1]{\multicolumn{15}{P{\lenc}|}{#1}}

\title{DeepWait: Pedestrian Wait Time Estimation in Mixed Traffic Conditions Using Deep Survival Analysis}

\author{
  Arash Kalatian \\
  Laboratory of Innovations in Transportation \\
  Ryerson University\\
  Toronto, ON \\
  \texttt{arash.kalatian@ryerson.ca} \\
   \And
 Bilal Farooq \\
  Laboratory of Innovations in Transportation\\
  Ryerson University\\
  Toronto, ON \\
  \texttt{bilal.farooq@ryerson.ca} \\
}

\begin{document}
\maketitle

\begin{abstract}
Pedestrian's road crossing behaviour is one of the important aspects of urban dynamics that will be affected by the introduction of autonomous vehicles. In this study we introduce DeepWait, a novel framework for estimating pedestrian's waiting time at unsignalized mid-block crosswalks in mixed traffic conditions. We exploit the strengths of deep learning in capturing the nonlinearities in the data and develop a cox proportional hazard model with a deep neural network as the log-risk function. An embedded feature selection algorithm for reducing data dimensionality and enhancing the interpretability of the network is also developed. We test our framework on a dataset collected from 160 participants using an immersive virtual reality environment. 
Validation results showed that with a C-index of 0.64 our proposed framework outperformed the standard cox proportional hazard-based model with a C-index of 0.58.  
\end{abstract}


\section{Introduction}
\label{S:1}

Extensive presence of autonomous vehicles (AVs) in urban areas is going to be one of the major changes in near future. AVs can fundamentally alter dynamics of urban areas. One of the aspects of urban mobility that can be affected by the widespread adaption of AVs is pedestrian's crossing behaviour. Studying this behaviour is important for the safety of most vulnerable users, optimal roadway layout changes, geometric design updates, and traffic flow optimization. AVs yielding to crossing pedestrians in a pedestrian-friendly urban area raises the necessity for new approaches in studying pedestrian-vehicle interactions~\cite{millard2018pedestrians}, which emphasizes on the advantages of mid-block unsignalized crossings. In an attempt to study pedestrians' crossing behaviour, in this paper we investigate the pedestrian waiting time before crossing mid-block unsignalized crosswalks in fully automated, mixed-traffic, and fully human-driven conditions. As it is impossible to collect real data for studies on future technologies and their impact, and using questionnaires and stated preferences surveys may result in unrealistic answers~\cite{farooqvire}, an immersive Virtual Reality (VR) based 3D environment was designed for the purpose of this study. Participants were asked to cross a simulated crosswalk in different scenarios, which were empowered by theoretical designs of experiments. In the meantime, participant's reactions and behaviours, , i.e. their coordinates, head orientations, stress level, etc. were being recorded, adding up to their sociodemographic data collected by questionnaires before the experiments.

Using the collected data, a cox proportional hazard model is developed to estimate pedestrian's waiting time based on different factors. To better capture nonlinearities involved in the data, we then proposed DeepWait, a deep neural network based cox proportional hazard model, backed up by a feature selection algorithm to select most important factors affecting pedestrian waiting time.

The rest of this paper is organized as follows: In the next section, an overview of the literature on pedestrian crossing behaviour, and particularly waiting time is provided. Next, details of the DeepWait modelling framework  are discussed. We then describe data collection procedure. Results and analysis of the cox proportional hazard model and deep survival analysis are then elaborated in detail in section \ref{S:5}. Finally, we present a summary of the work, as well as conclusions and some possible future directions of the study.

\section{Background}
\label{S:2}
Depending on the crossing area, studies on pedestrians crossing behaviour can be categorized into two main groups: 1. Crossing at signalized intersections, and 2. Crossing at unsignalized intersection or mid-blocks. Although there are wide variety of studies on pedestrian crossing behaviours, e.g. human-vehicle interactions~\cite{sun2003modeling,chen2016interaction}, pedestrian gap acceptance~\cite{das2005walk,oxley2005crossing}, pedestrian violations~\cite{keegan2003modifying}, etc.,  in this section we focus on studies explicitly concerning waiting time. 

In one of the earliest studies of pedestrian crossings at unsignalized crosswalks, Hamed~\cite{hamed2001analysis} used data collected from observations in the city of Amman, Jordan, to investigate the pedestrian waiting time at mid-block locations on divided and undivided roads. A cox proportional hazard model was developed to model pedestrian waiting time on the sidewalks. The results suggested that having accident experience, car ownership, number of people on the crosswalk, age, gender, type of trip and vehicle gap time are important factors in determining wait time before crossing. It is interesting to note that the parameters related to traffic were not found to be statistically significant. In another study, Tiwari~\textit{et al.}~\cite{tiwari2007survival}, observed that the average waiting time for females are 27\% more than for males at signalized intersections. In a similar work, Wang~\textit{et al.}~\cite{wang2011individual} developed a parametric survival analysis and found that personal characteristics affect crossing behaviour. In their sample group of study, young men on their way to school were more likely to terminate their waiting time in shorter time. They also indicated that crossing behaviour is highly influenced by safety awareness and conformity behaviour. In another study on modeling pedestrians' waiting time at signalized intersections, Li~\cite{li2013model} applied a statistical modeling approach and proposed a U shaped distribution for pedestrian waiting time at signalized intersections. By developing the model for different age cohorts, it was concluded that the waiting time increases with age.

The strength of survival analysis in modelling duration times in various fields, particularly in medical research, made us explore these approaches on our specific problem. We started with analyzing distracted pedestrian's waiting time using Cox Proportional Hazard models, and addressed various contributing distraction factors in~\cite{kalatian2018cox}. Here we develop a deeper investigation of waiting time in mixed traffic conditions and also contribute to the new survival analysis methods where neural networks embedding are exploited in the risk functions. Although the idea of incorporating neural networks in cox proportional hazard models is not new~\cite{faraggi1995neural}, only recently, and with the advances in computational powers and introduction of deep neural networks, researchers have successfully developed survival analysis models with better performances than traditional cox proportional hazard models~\cite{katzman2016deep,nezhad2019deep}. By implementing deep neural networks in hazard models, the underlying assumption of linearity in cox proportional hazard model is no longer necessary and one can observe the nonlinearities among complex data available today.   However, despite the successful implementations of deep neural networks in survival analysis in other fields, to the best of our knowledge no study in the field of transportation has yet adapted advanced survival analyses. Furthermore, our study develops a systematic way of selecting most useful covariates for the model. In the next section, a brief overview of the development of survival methods is provided. 
\section{DeepWait Framework}
\label{S:3}
\begin{figure}
  \centering
  \includegraphics{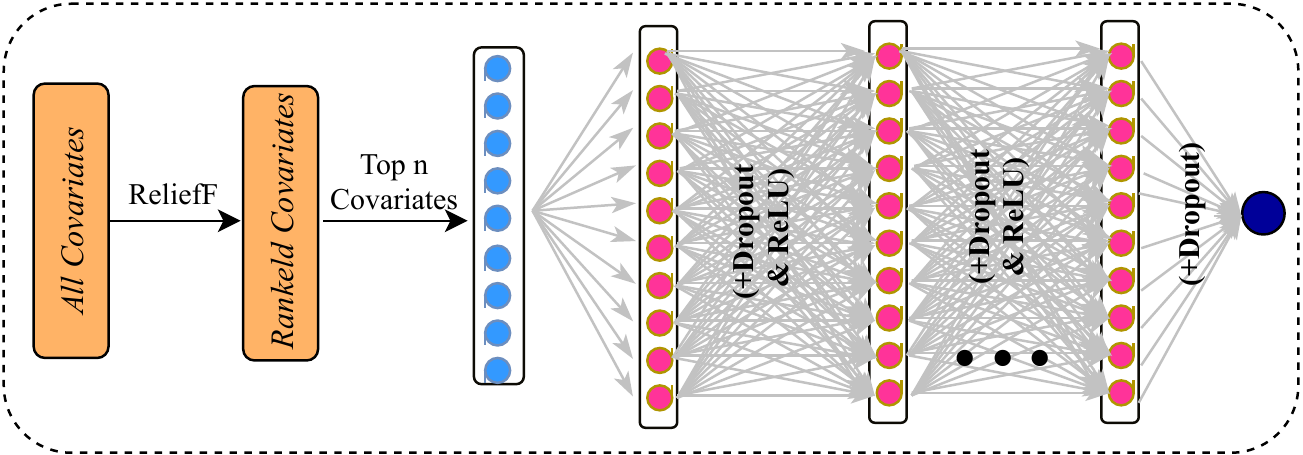}
  \caption{DeepWait framework}
  \label{fig:net}
\end{figure}

In safety analysis, survival methods have conventionally been the dominant method. In these methods, Hazard function, denoted by $\xi$(t), is defined as the probability of the occurrence of the event in an infinitesimal period of time ($\Delta t$) after time $t$. Equation (\ref{eq1}) provides the mathematical form of hazard function: 
\begin{equation}\label{eq1}
	\xi(t)=\lim_{\Delta t\to 0}\frac{Pr[t+\Delta t \ge T \ge t|T\ge t]}{\Delta t} 
\end{equation}
In the context of our study, an event is the initiation of the cross by participants. Thus, a greater hazard means a higher probability that the pedestrian starts crossing the street.\\  
Cox Proportional Hazard model (CPH) has been the most popular tool for analyzing survival data. Developed by Cox in 1972~\cite{cox1972regression}, Cox proportional-hazards model is essentially a regression model mainly used in medical research to identify the relationship between survival times of patients and predictor variables. As the base line model in this study, a CPH model with a linear log-risk function is developed:
\begin{equation} \label{eq2}
	\xi(t|R)=\overline{\xi}(t) \times e ^ {\displaystyle \sum_{i=1}^{n} \chi_i R_i}
\end{equation}
In this model, R is the vector of independent variables, $\chi$ is the vector of parameters to be estimated, and $\xi$(t) is the underlying hazard \cite{therneau2013modeling}. Equation (\ref{eq2}) gives the risk at time $t$ for pedestrian $i$, where the underlying hazard expresses hazard or risk for a pedestrian at all time points regardless of the independent variables (R=0). To estimate model parameters, partial derivatives of the log-likelihood function is taken. The underlying assumption in the CPH model is the linearity of log-risk function, i.e. ${ \sum_{i=1}^{n} \chi_i R_i}$. This assumptions may not always hold true, or may be too simplistic in some cases. Particularly, considering that these days complex data is available via advanced techniques and tools. To expand the power of the model and account for nonlinearities within data, in DeepWait framework, similar to the cox proportional hazard deep neural network proposed in \cite{katzman2016deep}, we replace the linear log-risk function in CPH with a deep neural network with a loss function proportional to the negative of partial likelihood of CPH. As depicted in Fig. \ref{fig:net}, the input layer for the network would be the layer of covariates, with multiple fully connected hidden layers, followed by dropout layers. In the end, the output would be a one-node layer estimating log-risk function.

However, the \textit{black box} nature of neural networks, i.e. not being able to look into the network and interpret the procedures happening inside it, makes \cite{katzman2016deep} and other similar methods less powerful than their linear regression-based counterparts in terms of interpreting results and finding the contributing covariates to waiting time. To tackle this major issue, DeepWait has Relief-based feature selection algorithm embedded within the deep CPH framework. The main application of Relief algorithms is to determine rank variables based on their importance. By removing irrelevant covariates, calibration process will become more efficient as non-useful predictors add to computational complexity and make interpretation of the results more difficult~\cite{robnik2003theoretical}. Proposed by Kononenko \textit{et al.} in 1997 for  classification \cite{kononenko1997overcoming}, Relief Algorithms family estimate the quality of variables based on their ability to distinguish between observations that are close to each other. By implementing RReliefF algorithm based on each of the newly created variables, importance weights can be calculated for each covariate. 

To train DeepWait framework, multiple hyper-parameters need to be tuned. These hyper-parameters include: number of hidden layers, number of nodes in each hidden layer,  learning rate, regularization coefficient, dropout rate, and exponential learning rate decay and momentum for Stochastic Gradient Descent (SGD) used in optimization. In addition, in each network, only top $n$ features based on their rank will be selected to train the network. $n$, which is also equal to the number of nodes in the input layer is the last hyper-parameter that is tuned in the network.

Finally, to evaluate the performance of the proposed models and compare them, the concordance-index (C-index)~\cite{harrell1984regression} is used as the most common method in survival analysis. C-index indicates the probability that a sample with higher calculated risk, experiences the event before a sample with lower calculated risk, where samples are selected randomly.

\section{Data Collection}
\label{S:4}
Virtual Immersive Reality Environment (VIRE)~\cite{farooqvire} was used in this experiment to simulate an unsignalized cross-walk with a mixed-traffic. Before each experiment, participants were asked to fill a form on their sociodemographic information, walking habits, health conditions, and previous VR experiments. Table \ref{tab:Tripsum} shows various levels of parameters that define each scenario. For variables related to regulations and standards, i.e. speed limit, lane width and minimum gap, one level is set to be the current standard of urban areas in Toronto. Levels of flow rate are assigned similar to three congestion levels on a similar downtown Toronto intersection. Number of breaking levels, which is defined only for autonomous vehicles, defines the smoothness of the break of AVs when confronting pedestrians. More levels mean a smoother break with more levels on deceleration. 

After the experiments, based on the reactions of participants, their Post Encroachment Time (PET) and distance to collision were also calculated as surrogate safety measures. Finally, the time participants wait on the virtual sidewalk before initiating a cross is calculated as the dependent variable in the study.

\begin{figure}[ht]
\centering
\begin{subfigure}{0.39\textwidth}
    \centering
    \includegraphics[width=0.7\textwidth,height=6cm]{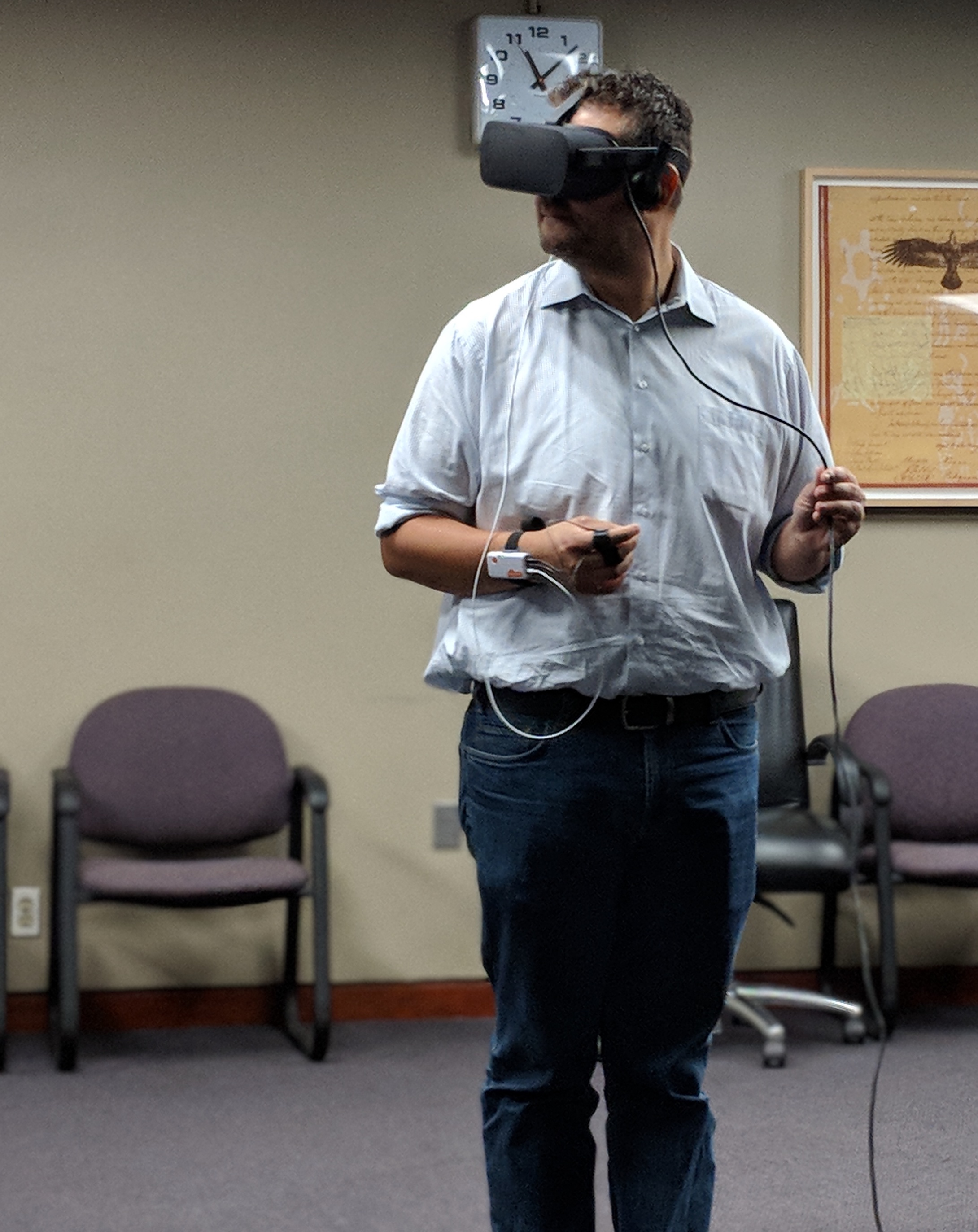}
    \caption{A participant during the experiment}

\end{subfigure}
\begin{subfigure}{0.6\textwidth}
    \centering
    \includegraphics[width=0.9\textwidth,height=6cm]{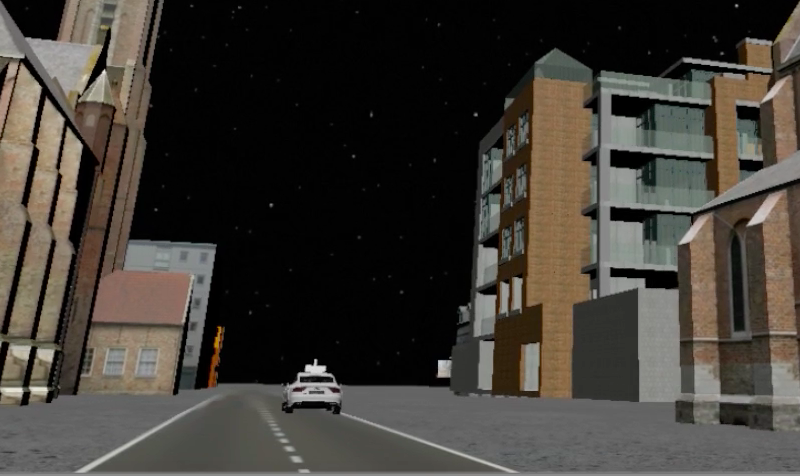}
    \caption{A sample environment}
\end{subfigure}
\caption{Experiment setup}
\label{fig:vire}
\end{figure}

Data collection campaign was conducted during summer 2018, in four different locations in Greater Toronto Area to account for heterogeneity of the population. Fig. \ref{fig:vire} depicts a participant while doing the experiment and a sample scenario that participants are exposed to. A total of 160 people participated in the experiments, including 113 adults and 47 kids. Each adult was exposed to 15 scenarios, repeated twice to make a 30-scenario experiment. In this particular study, data collected from kids are not taken into account as their crossings, when exposed to virtual autonomous vehicles, involved unpredicted reactions and required further data cleaning procedures.

After removing experiments where pedestrians could not complete the crossing, or could not follow the experiments' procedures, a total of 2463 responses were remained to be studied. Collected data is then used to estimate models for predicting pedestrian waiting time.
\begin{table*}
\footnotesize
\caption{\footnotesize Levels of Adjustable Parameters}
\setlength\extrarowheight{2pt}
\centering

\begin{tabular}{|l|*{30}{p{\lena}|}}
\hline
\bf{Variable}           & \multicolumn{30}{c|}{\bf{Levels}}\\
\hline\hline  
Speed limit (km/h) & \mcii{30} & \mcii{40} & \mcii{50} \\
\hline
Lane width (m) & \mcii{2.5} & \mcii{2.75} & \mcii{3} \\  
\hline
No. of breaking levels & \mcii{1} & \mcii{2} & \mcii{3} \\  
\hline
Minimum Gap (s) & \mcii{1} & \mcii{1.5} & \mcii{2} \\  
\hline
Flow rate (veh/hr) & \mcii{530} & \mcii{750} & \mcii{1100} \\  
\hline
Road type & \mcii{1-way} & \mcii{2-way} & \mcii{2-way \& median} \\  
\hline
Traffic Automation Status& \mcii{Human Driven} & \mcii{Mixed Traffic} & \mcii{Fully Automated} \\  
\hline
Time of day & \mciii{Day} & \mciii{Night}  \\
\hline
Weather & \mciii{Clear} & \mciii{Snowy}  \\
\hline
\end{tabular}
\label{tab:Tripsum}
\end{table*}

\section{Results}
\label{S:5}
In Table \ref{tab:mean}, average waiting time of participants for different levels of covariates are compared. As it can be observed in this table, females in our sample data have longer waiting times compared to males. As for the age of participants, pedestrians older than 50 have longest waiting times among all other age brackets. Having a previous experience with the VR equipment makes the pedestrians feel more comfortable, thus leading to a shorter stay on the sidewalk. 

Comparing waiting time of different variables w.r.t. mobility habits reveals that people in our sample who tend to walk more, are more cautious crossing the street and hence have longer waiting times.

Regarding scenario parameters, the average waiting time for different speed limits does not follow the commonsense expectation. Based on our sample data, participants tend to have shorter waiting time in faster traffics. This can be explained by the observation that participants in our experiment were more inclined to wait for the street to be totally cleared, which takes more time in slow moving traffics. Replacing speed limit with variables such as accepted and rejected gap times can help address this issue in future studies. Going through the table, it can be inferred that pedestrians wait longer before crossing a lane with greater width.

Having a 1 second allowable gap time has led to a longer waiting time for participants in our study. Waiting time for different arrival rates of the simulated vehicles shows that pedestrians in our experiments have longer waiting times in more congested areas. Results show that in our sample data, fully autonomous scenarios lead to longer waiting times, compared to human-driven scenarios. However, mixed traffic conditions with both human-driven and autonomous vehicles leads to longest waiting times, probably as a result of confusion of participants over different types of cars and their capabilities. Next variable in the table is the braking level, which is defined as three different ways of applying deceleration. However, it seems that the effect of different levels of braking systems on wait time do not follow a particular trend, which can be traced back to the fact that the participants were not aware of the definitions of braking systems in the experiment. Road type specifications reveal that having a median in the experiment makes participants wait shorter on the sidewalk, compared to two-way roads. Finally, participants in snowy weather conditions waited longer on the sidewalk, whereas waiting time on day and night in our experiment appears to be similar to each other. 

\begin{figure}

    \centering
    \includegraphics[scale=.6]{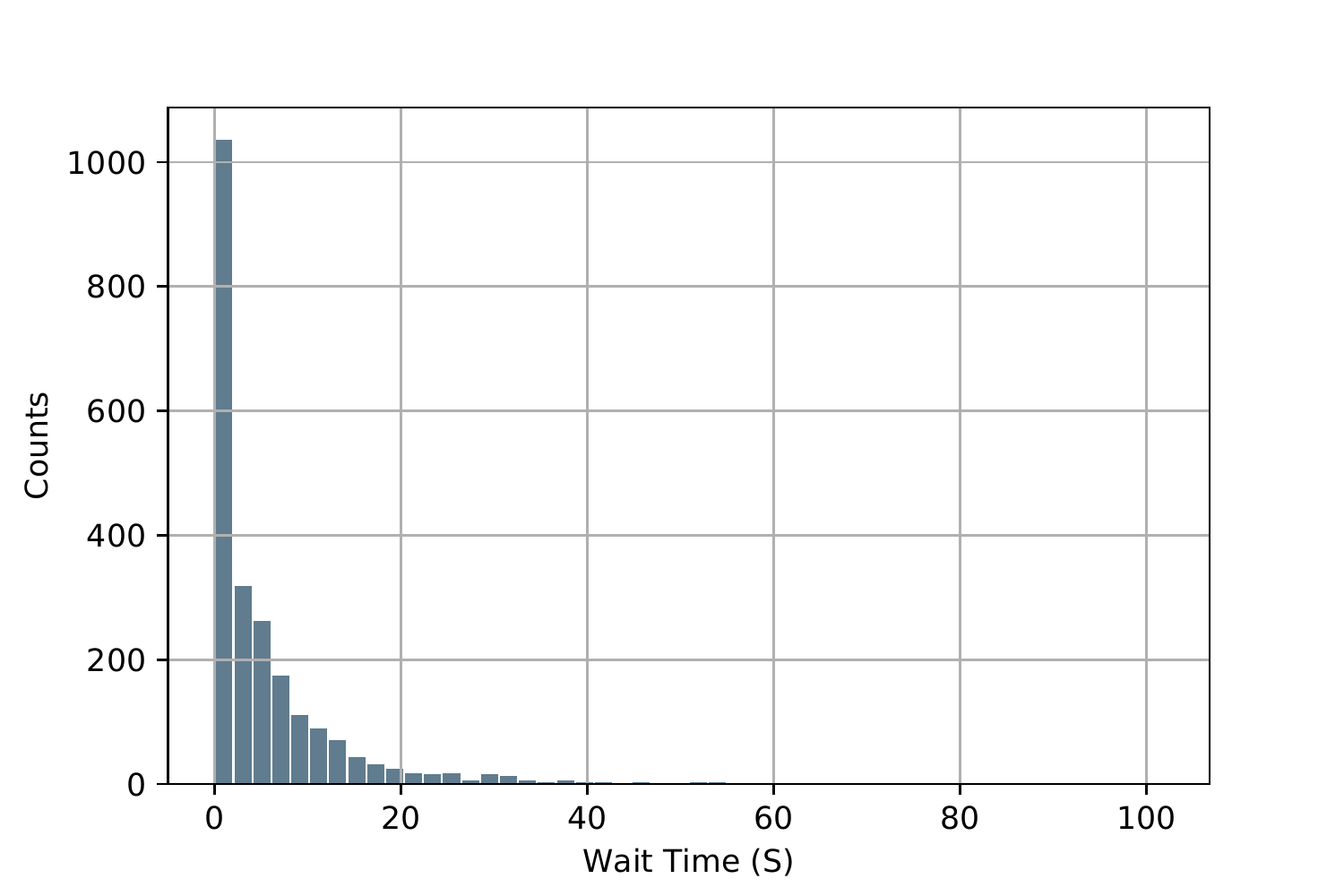}
    \caption{Waiting Time frequency distribution}
    \label{fig:waitfreq}
\end{figure}

Fig. \ref{fig:waitfreq} presents the frequency distribution of waiting time in experiments. As it can be seen in this figure, more than 50\% of crossings in our experiment occur in the fist two seconds.

\begin{table}
\caption{Average Waiting time for different levels}  
\footnotesize
\begin{center}
\begin{tabular}{|c|c|c|}
  \hline 
  \textbf{Variable} & \textbf{Level} & \textbf{Wait Time (s)}  \\
  \hline
  \multicolumn{3}{|c|}{\textit{Sociodemographic}}\\
  \hline
  Gender & \begin{tabular}{@{}c@{}}Male \\ Female\end{tabular} & \begin{tabular}{@{}c@{}}5.4 \\ 6.6\end{tabular}  \\
  \hline
  Age & \begin{tabular}{@{}c@{}}18-24 \\ 25-29\\30-39 \\40-49 \\ over 50\\ \end{tabular} & 
    \begin{tabular}{@{}c@{}}7.4 \\ 5.6\\4.3 \\4.6 \\ 9.1\\ \end{tabular}\\
  \hline
  Previous VR experience & \begin{tabular}{@{}c@{}}Yes \\ No\end{tabular} & \begin{tabular}{@{}c@{}}5.8 \\ 6.0\end{tabular}  \\
  \hline
    \multicolumn{3}{|c|}{\textit{Transportation Information}}\\
  \hline
  Driving Licence & \begin{tabular}{@{}c@{}}Yes \\ No\end{tabular} & \begin{tabular}{@{}c@{}}5.9 \\ 6.8\end{tabular}  \\
  \hline
  Maid Mode of Transport & \begin{tabular}{@{}c@{}}Bike \\ Car\\Public Transit \\Walking \\ \end{tabular} & 
    \begin{tabular}{@{}c@{}}5.0 \\ 7.2\\4.7 \\7.5 \\ \end{tabular}\\
  \hline
  Walk to Work & \begin{tabular}{@{}c@{}}Yes \\Sometimes\\ No\\\end{tabular} & \begin{tabular}{@{}c@{}}8.1 \\ 5.1\\5.6\\\end{tabular}  \\
  \hline
  \multicolumn{3}{|c|}{\textit{Experiment Factors}}\\
  \hline
  Speed Limit & \begin{tabular}{@{}c@{}}30 \\ 40\\50\end{tabular} & \begin{tabular}{@{}c@{}}7.1\\5.7\\5.0\end{tabular}  \\
  \hline
  Lane Width & \begin{tabular}{@{}c@{}}2.5 \\ 2.75\\3 \\ \end{tabular} & 
    \begin{tabular}{@{}c@{}}4.8 \\ 6.5\\6.6\\ \end{tabular}\\
  \hline
  Minimum Gap & \begin{tabular}{@{}c@{}}1 \\1.5\\ 2\end{tabular} & \begin{tabular}{@{}c@{}}6.3 \\ 5.4\\6.1\end{tabular}  \\
  \hline
    Mean Arrival Rate & \begin{tabular}{@{}c@{}}530 \\ 750\\1100\end{tabular} & \begin{tabular}{@{}c@{}}4.3\\5.3\\8.4\end{tabular}  \\
  \hline
  Traffic Automation Status\% &
  \begin{tabular}{@{}c@{}}Human Driven \\ Mixed Traffic\\Fully Automated\end{tabular} & \begin{tabular}{@{}c@{}}4.9\\6.9\\5.9\end{tabular}  \\
  \hline
  Braking Levels & \begin{tabular}{@{}c@{}}1 \\ 2\\3 \\ \end{tabular} & 
    \begin{tabular}{@{}c@{}}5.7 \\ 6.2\\6.0\\ \end{tabular}\\
  \hline
  Road Type & \begin{tabular}{@{}c@{}}One-Way \\Two-Way\\ Two-Way with median\end{tabular} & \begin{tabular}{@{}c@{}}6.7 \\ 6.2\\4.8\end{tabular}  \\
  \hline
      Weather Condition & \begin{tabular}{@{}c@{}}Clear \\ Snowy\end{tabular} & \begin{tabular}{@{}c@{}}5.7\\6.3\end{tabular}  \\
  \hline
        Time & \begin{tabular}{@{}c@{}}Day \\ Night\end{tabular} & \begin{tabular}{@{}c@{}}5.9\\5.8\end{tabular}  \\
  \hline

\end{tabular}
\label{tab:mean}
\end{center}
\end{table}

\subsection{Cox Proportional Hazard Model}
To develop a cox proportional hazard model, the first step would be to remove highly correlated variables which would lead the convergence to be halted due to singular matrices not having an inverse. In this study, we used variance inflation factor (VIF) to spot redundant variables and remove them. In the next step, model specifications were determined through a systematic process of removing statistically insignificant variables. 
Table \ref{tab:cox} presents the results of the cox proportional hazard model with linear assumption in log-risk function. The first column in this table shows covariates that appeared to be significant (95\% confidence level). It should be noted that a positive coefficient sign means that the hazard (chance of starting a crossing in this case) is higher, and thus waiting time is shorter. The hazard ratios (exponential of coefficients) greater than one indicates that as the value of the variable increases, the event hazard increases and thus waiting time decreases.

As it can be seen in Table \ref{tab:cox}, speed limit is one of the significant factors in our experiment with a positive coefficient, meaning that participant waited longer in our experiment in slower traffics. As discussed in the previous section, participants in our experiment waited for the street to be totally cleared, which takes more time in slow moving traffics. Further analysis in our future studies is required on this side to account for this issue. Lane width is the other contributing factor in the CPH model. Participants waited for shorter times on narrower simulated lane widths. Arrival rate of the cars in the traffic has a significant effect on users' waiting times. Negative coefficient of this variable shows the fact that in higher congestion, it is more difficult for pedestrians to find a safe gap to cross the street. Results also reveal that by adding a median to the simulated crosswalk, pedestrians waiting time can be decreased.  Age is the other contributing factor in waiting time. The negative value for the variables shows that waiting time for participants in oldest and youngest age brackets is generally longer than that of other participants.  Additionally, results show that pedestrians who do not walk for doing errands, waited longer on the sidewalk. Moreover, participants with the main mode of public transit have shorter waiting times. Previous experience with VR equipment also appears to be a significant factor in determining waiting time. Despite allocating 2 trial experiments for each participant to get adapted with the environment, participants with no previous VR experiment required more time to cross the simulated street. This factor can be considered in the future experiments to allocate more time for training participants. Finally, female pedestrians have longer waiting times. 
\begin{table}[t]
\footnotesize
\caption{Multivariate proportional hazard model results}
    \centering
    
    \begin{tabular}{|c|c|c|c|}
    \hline
         \textbf{Variable}&\textbf{Coefficient}&\textbf{Hazard Ratio}& \textbf{p-value}  \\
    \hline
         Speed Limit&0.01&1.01&$<$0.005\\
    \hline
         Lane Width&-0.59 & 0.55&$<$0.005\\
    \hline
         Mean Arrival Rate&-0.01& 1.01&$<$0.005\\
    \hline
         Road Type: Two-way with Median&0.19& 1.21&$<$0.005\\
    \hline
         Age: Over 50&-0.43& 0.65&$<$0.005\\
    \hline 
         Age: 18-24&-0.27&0.76&$<$0.05\\
    \hline
         Main Mode: Public Transit&0.23   &   1.26&$<$0.005\\
    \hline
         Walk to shop: No&-0.19&  0.83&$<$0.005\\
    \hline
         Previous VR Experience: Yes&0.19&  1.21&$<$0.005\\
    \hline
         Gender: Female&-0.11& 0.90&0.02\\
    \hline
    
    \end{tabular}
    
    \label{tab:cox}
\end{table}
Using 10-fold cross-validation, the average C-index of the final model appears to be 0.58.
\subsection{DeepWait Framework Analysis}
First we developed a deep neural networks (DNN) using all covariates as features. The C-index achieved is 0.61, which is 3 percent greater than that of linear CPH model. The difference is close to what was achieved in the original paper of by \cite{katzman2016deep} their datasets i.e. Support, METABRIC, and Rotterdam \& GBSG.

The first step towards our DeepWait framework is to rank features based on their importance using RReliefF algorithm. We then introduce hyper-parameter $n$ in the deep network as the number of variables to be selected, i.e. the top $n$ variables based on their importance will be selected for the deep neural network. To tune the network, a random search for hyper-parameter optimization~\cite{bergstra2012random} is ran and 100 neural networks are trained to find the optimum network for the collected data. To evaluate the performance of the networks, 10-fold cross-validation is conducted and highest C-index on validation set is reported as the best performance of the framework. Based on the performance of 100 neural networks over 1000 epochs, a framework with 17 top features as inputs, three hidden layers each with 75 nodes, no batch normalization and a dropout rate of 0.1 outperformed all the other models with a C-index of 0.64. Top 17 features selected in this framework along with their average ReliefF importance weights are presented in table \ref{tab:rank}. Comparing this table with table~\ref{tab:cox}, we can see that 7 variables, including mean arrival rate, previous VR experiment, no walking for shopping, having medians in two-way roads, and age and gender variables are repeated covariates in both models. On the other hand, variables related to speed limit and lane width, and main mode of public transit do not appear to be a part of the contributing factors in the proposed framework. Elimination of main mode of public transit may be traced back to the inclusion of two other main modes of transport in DeepWait model. Speed limit and lane width variables are also removed in the availability of other variables that appeared to be important considering nonlinearities. Finally, another age category variable, one environmental variables (weather condition), mixed traffic condition, education and employment variables, and number of cars in the household appear to be contributing to our proposed model. As it can be inferred from the results, the nonlinear survival model could capture the effect of mixed traffic conditions on waiting time. To further quantitatively investigate the effects of these covariates on pedestrian waiting time, advanced interpretability procedures are required, which is beyond the scope of this paper and will be investigated in future works. 

Table~\ref{tab:compare} compares C-index for three models and number of features used. As provided in this table, using our framework, we successfully improved the accuracy of linear CPH by 6\%, and DNN by 3\%. Comparing to CPH, our model does not require hand-crafted features and various manual model developments to find the optimum model. Moreover, our framework uses less features, and results in a higher accuracy with a less computationally expensive solution, compared to the use of DNN. 
\begin{table}[t]
\footnotesize
\caption{Selected features for DeepWait network}

\centering
\begin{tabular}{|c|c|c|}
\hline
   \textbf{Rank}&\textbf{Covariate}  & \textbf{Importance Weight} \\
\hline
     1& Walk for Shopping: No &0.029\\

     2&Education: High School Diploma & 0.025\\
     3&Age: over 50&0.022\\
     4&Main Mode: Car  &0.019\\
     5& Walk to work: No &0.019\\
     6&Mean Arrival Rate&0.019\\
     7&Education: Masters Degree & 0.017\\
     8&Weather: Snowy&0.011\\
     9&Age: 18-24&0.011\\
     10&Occupation: Unemployed&0.011\\
     11&Number of Cars in the Household&0.009\\
     12&Gender: Female&0.009\\
     13&Main Mode: Walk&0.008\\
     14&Road Type: Two-way with Median&0.003\\
     15&Previous VR experience: Yes&0.003\\
     16&Traffic: Mixed&0.002\\
     17&Age: 25-29  &0.002\\
     \hline
\end{tabular}
\label{tab:rank}
\end{table}
\begin{table}[t]
\footnotesize
\caption{Comparing the performance of the models}

\centering
\begin{tabular}{|c|c|c|}
\hline
   \textbf{Model}&\textbf{Number of Covariates}  & \textbf{C-index} \\
\hline
     Linear CPH&10 &0.58\\
     DNN CPH&47& 0.61 \\
     DeepWait&17 & \textbf{0.64}\\
     \hline
\end{tabular}
\label{tab:compare}
\end{table}
\section{Summary and Conclusions}
\label{S:6}
In this paper, we studied one of the main aspects of pedestrian crossing behaviour, i.e. waiting time at unsignalized crosswalks. With the growing interest in autonomous vehicles, human-vehicle interactions will change dramatically considering that AVs can detect and yield to pedestrians. To investigate this issue, we developed a virtual reality-based data collection campaign in Toronto, and asked 160 participants to cross an unsignalized mid-block crosswalk in mixed traffic conditions, while their reactions were being tracked and recorded. Our results can help city planners and decision makers better prepare urban areas and and propose regulations for the future. In our study, results from linear CPH model reveals that roads with narrower lane widths or medians help pedestrians feel more comfortable and have shorter waiting times. City planners can consider these changes to update urban areas to account for such behaviours. Moreover, areas with older demographics may require more attention in terms of pedestrians crossings.   \\
To estimate pedestrian waiting time, we proposed DeepWait, a deep neural network cox proportional hazard model with embedded feature selection. The framework proposed in this study enables a more accurate estimation of waiting time for pedestrians by incorporating the strength of deep neural networks within the cox proportional hazard models. Deep networks enable us to capture the nonlinearities better, which is in particular beneficial for high dimensional data made available by novel data sources. Moreover, by implementing and following a systematic feature selection algorithm within our framework, our proposed method does not require manual selection of significant covariates as in traditional CPH. Moreover, unlike \cite{katzman2016deep} that uses all the covariates as input features, by selecting more important covariates irrelevant features are removed and the accuracy of the model improves by 3\%. Selecting features also enhance the interpretability of the model, by introducing more contributing covariates.

Although feature selection algorithm helps us with recognizing more contributing covariates, the systematic sensitivity analysis of the waiting time to these covariates will be investigated in the future. This issue will be addressed by implementing interpretability techniques that have been emerging in recent years. In terms of the model, more advanced deep learning networks, e.g. residual networks and convolutional neural networks, can be implemented and compared in future studies. Finally, more features can be derived and used from the collected data to estimate waiting time based on a wider set of covariates.

\bibliographystyle{unsrt}  


\end{document}